\documentclass[12pt, a4paper]{article}
\usepackage{latexsym,amsmath,amsfonts,amssymb}
\usepackage{graphicx}
\usepackage{indentfirst}
\usepackage{cite}
\usepackage[cp1251]{inputenc}

\newtheorem{theorem}{Theorem}
\newtheorem{lemma}{Lemma}

\newtheorem{remark}{Remark}

\newcommand{\ba}{\begin{array}}
\newcommand{\ea}{\end{array}}
\newcommand{\be}{\begin{equation}}
\newcommand{\ee}{\end{equation}}

\begin{document}

\begin{center}
 {\Large \bf Lie and conditional symmetries of a class of nonlinear  (1+2)-dimensional boundary value problems}

\medskip

{\bf Roman Cherniha~$^{a b}$  and John R King~$^a$}
\\
\medskip
{\it $^a$ School of Mathematical Sciences,
University of Nottingham, \\
University Park, Nottingham, NG7 2RD, UK}\\
{\it $^b$~Institute of Mathematics,  Ukrainian National Academy
of Sciences,\\ 3 Tereshchenkivs'ka Street, 01601 Kyiv, Ukraine}
\\
\medskip
 E-mail: cherniha@gmail.com; John.King@nottingham.ac.uk
\end{center}

\begin{abstract}

 A new  definition  of conditional invariance for boundary value problems involving a wide range of boundary conditions
 (including initial value problems as a special case)  is proposed.  It is shown that    other definitions
  worked out    in order to find Lie symmetries  of   boundary value problems  with standard boundary
  conditions,   follow as    particular cases  from our definition.  Simple examples of  direct  applicability
    to the nonlinear   problems  arising in applications are   demonstrated.
 Moreover, the  successful application  of the
  definition for the Lie  and conditional symmetry classification of
 a  class of (1+2)-dimensional nonlinear boundary value problems   governed by the  nonlinear diffusion equation
  in a semi-infinite
 domain is realised. In particular, it  is  proved that there is a special exponent, $k=-2$,  for the power
 diffusivity $u^k$ when the problem in question  with non-vanishing flux on the  boundary admits additional Lie symmetry
 operators  compared to the case $k\not=-2$.
In order to demonstrate the applicability of the symmetries derived,
 they are  used   for reducing
  the nonlinear problems  with power diffusivity $u^k$ and a constant non-zero flux on the boundary
   (such problems are    common in applications and describing a wide range of phenomena) to
     (1+1)-dimensional  problems. The structure and  properties of the  problems obtained   are briefly analysed.
     Finally, some results demonstrating how   Lie  invariance of the boundary value
     problem in question depends on geometry of  the domain are presented.

\end{abstract}



\medskip


\section{Introduction }

Nowadays  the Lie symmetry method
 is widely applied
to study partial differential equations (including multi-component systems
of multidimensional PDEs), notably  for  their reductions to ordinary differential equations (ODEs) and
for constructing  exact solutions.
There are a huge number of papers and many excellent
books (see, e.g.,  \cite{bl-anco02, bl-k, fss, olv, ovs} and papers
cited therein) devoted to such applications.
During recent decades,  other symmetry  methods, which are  based on  the   classical Lie  method, were  derived.
 The  Bluman-Cole  method of non-classical  symmetry (another widely used terminology is Q-conditional symmetry,
  proposed in \cite{fss}) is  perhaps the best known   among them  and  the recent  book  \cite{bl-anco-10} summarizes
   results  obtained  by means of this approach for scalar PDEs (see also  the recent
papers \cite{ch-2010, ch-dav-2013} for some results and references
in the case of nonlinear PDE systems).

However, a PDE   cannot  model any real process without additional condition(s)  on the unknown function(s),
 a boundary value problem (BVP)  based on the given PDE   being needed to  describe  real processes arising
 in nature or society. One may note that  symmetry-based methods  have not  been   widely used   for solving   BVPs
(we include initial value  problems
within this terminology defining the initial condition as a
particular   case of a boundary condition). The obvious reason
follows from the following observation:   the relevant boundary and
initial conditions
 are usually not invariant under any transformations, i.e. they do not  admit any symmetry of the governing  PDE.
Nevertheless, there are some classes of BVPs  that  can be solved
   by means of  the  Lie   symmetry based     algorithm.  This algorithm uses the notion of   Lie invariance of
    BVP in question. Probably  the  first rigorous  definition of  Lie invariance for BVPs was formulated
     by  G.W. Bluman in the 1970s \cite{bl-1974} (the  definition and  several  examples  are  summarized
      in the book \cite{bl-anco02}). This definition was used (explicitly or implicitly) in several papers
      to derive exact solutions of  some BVPs. It should be noted that  Ibragimov's  definition of BVP
      invariance  \cite{ibr92} (see also his  recent paper \cite{ibr-11}),  which was formulated independently,
      is equivalent to Bluman's.
On the other hand, one notes that  Bluman's definition does not suit  all types of  boundary conditions.
Notably,
 the definition  does not  work   in the case of  boundary conditions  involving
     points at infinity.
        In  recent papers \cite{ch-kov11a,ch-kov11b,ch-kov12a},   a new definition   of   Lie invariance of
        BVPs
          with a wide range of  boundary   conditions (including those involving points at infinity and  moving surfaces)
             was  formulated. Moreover, an  algorithm for  the group classification for the given  class
             of BVPs was worked out  and   applied  to a class of  nonlinear two-dimensional and multidimensional
             BVPs of Stefan type  with the aim of  showing  their  efficiency.

     However, there are  many realistic BVPs that cannot be solved using any  definition   of   Lie
     invariance of  BVP, for instance   because  the relevant governing equations do not admit any Lie symmetry
      (or possess  a trivial one only).
     Hence, definitions involving  more general types of symmetries should be worked out. Having this in mind,
in this paper we consider
 a class of (1+2)-dimensional nonlinear boundary value problems (BVPs) modelling  heat transfer (for example)
  in the semi-infinite
 domain  $\Omega = \{(x_1, x_2): - \infty < x_1 < + \infty, \, x_2 > 0 \}$:
\begin{eqnarray}
& & \frac{\partial u}{\partial t} = \nabla.\left(d(u)\nabla u \right), \ \ (x_1, x_2) \in \Omega, \, t \in \mathbb{R}, \label{0.1} \\
& & \quad x_2 = 0: d(u)\frac{\partial u}{\partial x_2} = q(t), \label{0.2} \\
& & \quad x_2 \rightarrow + \infty:  \frac{\partial u}{\partial x_2}= 0 , \label{0.3}
\end{eqnarray}
where $u(t,x_1,x_2)$ is an unknown function describing a temperature field (say), $d(u)$ is the positive
coefficient of thermal conductivity,
 $q(t)$ is a specified  function describing the heat flux of energy absorbed at  (or radiating from)
  the surface $x_2 = 0$,  zero flux is prescribed at infinity
   (actually, one should use the condition
   $ d(u)\frac{\partial u}{\partial x_2}=0$ but we assume
   that two are equivalent provided $d(u)\not= 0$ when $x_2 \rightarrow + \infty$ ; a discussion of the dependence of the admissibility of such boundary conditions
   at infinity on $d(u)$ is a delicate one that lies outside the scope of the current work, however some results are presented in the end of Sections 3  and 4)
  and  the standard  notation $\nabla = \left(\frac{\partial}{\partial x_1}, \frac{\partial}{\partial x_2} \right)$
  is used.  Hereafter  we  assume that $d(u) \neq \mbox{constant}$ (otherwise the problem is linear and can
  be solved by the well-known classical methods) and  all the functions arising in problem (\ref{0.1})--(\ref{0.3})
  are sufficiently smooth. It should be noted that we do not prescribe any initial condition
  assuming that the initial profile can be an arbitrary smooth function that  can be specified to respect
  a symmetry of   BVP (\ref{0.1})--(\ref{0.3}) in question.

  \medskip

  The paper is organised as follows.  In Section 2,  a theoretical background is developed and  relevant
  examples are presented.
  In Section 3, the Lie symmetry classification of  BVPs of the form  (\ref{0.1})--(\ref{0.3}) is derived
  and the main result is presented in Theorem 2.  In Section 4, all possible reductions of  BVP   (\ref{0.1})--(\ref{0.3})
   with the power-law thermal conductivity $u^k$ and a non-zero constant flux $q(t)=q_0$
   that admit reductions to (1+1)-dimensional BVPs  are constructed. In Section 5, the  conditional  symmetry
   classification of the BVPs class  (\ref{0.1})--(\ref{0.3}) is derived and the relevant reductions
   are presented.  In Section 6, some results demonstrating how   Lie  invariance of BVP  in question
    depends on geometry of  the domain are presented.
    Finally, we  discuss the result obtained and  present some conclusions  in
the last section.

\section{Theoretical background:  definitions and examples }

Here we restrict ourselves to the case when the basic equation of
BVP is a multidimensional  evolution PDE of $k$th-order  in space
($k\geq 2$), i.e. our considerations here go well beyond the
specific equation  (\ref{0.1}).  Thus,  the relevant BVP may be
formulated as follows:
\begin{equation}
u_t=F\left(t, x, u, u_x, \ldots, u_{x}^{(k)}\right), \ x \in \Omega
\subset \mathbb{R}^n, \ t>0 \label{2-1}
\end{equation}
\begin{equation}
s_a(t,x)=0: \ B_a \left(t,x, u, u_x, \ldots, u_{x}^{(k_a)}\right) =
0, \ a = 1, 2, \ldots, p, \  k_a<k \label{2-2}
\end{equation}
where $F$ and  $B_a$ are smooth functions in the corresponding
domains,   $\Omega$  is    a domain with smooth
boundaries and  $s_a(t,x)$  are  smooth curves. Hereafter the subscripts
$t$ and $x = \left(x_1, \ldots, x_n \right)$, denote differentiation with respect to these variables and $u_{x}^{(j)}, \ j=1,\dots, k$ denotes a totality of partial derivatives of the order $j$  with respect to
the space variables, for example
$u^{(k)}_{x}=({u}_{x_1\dots x_1}, \dots, u_{x_{j_1}, \dots,x_{j_n}}, \dots, {u}_{x_n\dots x_n})$,  where  $u_{x_{j_1}, \dots,x_{j_n}} =  \frac{\partial^k u}{\partial x_{j_1} \ldots \partial x_{j_n}}, \, j = 1,2, \ldots,k; \, j_1+ \ldots+ j_n=k. $
 We assume that BVP (\ref{2-1})--(\ref{2-2}) has a classical
solution (in a usual sense).

 Consider the infinitesimal generator
\begin{equation}
X = \xi^0 (t,x)\frac{\partial}{\partial t}+\xi^i
(t,x)\frac{\partial}{\partial x_i} + \eta
(t,x,u)\frac{\partial}{\partial u}.\label{2-3}
\end{equation}
Hereafter  $\xi^0, \xi^i $ and $\eta$ are known smooth functions  and  summation is assumed from 1 to $n$ over repeated index $i$ in operators.  Assuming  that this operator
 defines a Lie symmetry acting  both on the $(t,x,u)$--space
and  on its projection to $(t,x)$--space, consider the  operator
  \be\label{2-3*}
  \medskip
   \mbox{\raisebox{-1.1ex}{$\stackrel{\displaystyle  
X}{\scriptstyle k}$}} =\mbox{\raisebox{-1.1ex}{$\stackrel{\displaystyle
X}{\scriptstyle k-1}$}}+\sigma^{k}_{1}\frac{\partial}{\partial{u}_{x_1\dots x_1}}
+\sigma^{k}_{2}\frac{\partial}{\partial{u}_{x_1\dots x_1x_2}}+\ldots+\sigma^{k}_{k_n}\frac{\partial}{\partial{u}_{x_n\dots x_n}},  \quad  k \ge 2
\ee
 corresponding to the
$k$th  prolongation of  $X$, whose   coefficients  
are calculated   via the functions  $\xi^0, \ldots, \xi^1, \eta $
and their derivatives by the well-known prolongation formulae
\cite{ovs, olv}  starting from the first prolongation of  $X$:
\[ \mbox{\raisebox{-1.1ex}{$\stackrel{\displaystyle  
X}{\scriptstyle 1}$}} = X  +\sigma^{1}_{0}\frac{\partial}{\partial
{u_{t}}} +\sigma^{1}_{1}\frac{\partial}{\partial {u_{x_1}}} +
\ldots+\sigma^{1}_{n}\frac{\partial}{\partial {u_{x_n}}}. \]
 In
formula   (\ref{2-3*}), $k_n$ is the total  number of different
$k$-order derivatives of the function $u$ w.r.t. the space variables
(there is no  need to take into account $k$-order derivatives  involving the time variable because
   (\ref{2-1}) contains the first-order time derivative only).

\noindent {\textbf{ Definition 1.\cite{bl-anco02}}}
The Lie symmetry $X$
 (\ref{2-3}) is admitted by the boundary value problem
 (\ref{2-1})--(\ref{2-2}) if
\begin{itemize}
\item[(a)] $
\mbox{\raisebox{-1.6ex}{$\stackrel{\displaystyle  
X}{\scriptstyle k}$}} \left(F\left(x, u, u_x, \ldots , u_{x}^{(k)}\right)
-u_t \right)=  0 $ when $u$ satisfies (\ref{2-1});
\item[(b)] $X (s_a(t,x)) = 0$ when $s_a(t,x) = 0, \ a = 1, \ldots,p$;
\item[(c)] $
\mbox{\raisebox{-1.6ex}{$\stackrel{\displaystyle  
X}{\scriptstyle k_a}$}} \left(B_a \left(t,x, u, u_x, \ldots ,
u_{x}^{(k_a)}\right) \right) =  0 $ when $s_a(t,x) = 0$  and  $B_a \vert_{ s_a(t,x) = 0}=
0$, \ $a = 1, \ldots,p$.
\end{itemize}

Because BVP  (\ref{2-1})--(\ref{2-2}) involves only the standard boundary conditions, Definition 1 cannot
 be applied to BVP (\ref{0.1})--(\ref{0.3}), which involves
boundary conditions  defined at  infinity.
Moreover,  Definition 1 cannot be generalised in a straightforward way to  the boundary condition
 (\ref{0.3}) (see an example in  \cite{ch-kov11b}).
This issue  was pointed out in \cite{king91}, where  it was    suggested  that an  appropriate substitution
be made  to  transform
   the unbounded domain to a bounded one.
   This idea was formalised in \cite{ch-kov11a, ch-kov11b}, where it was shown how this definition  can be   extended
to  classes of BVPs with more complicated boundary and initial conditions.
Here we go essentially  further, namely  we extend notion of BVP
invariance to  the case of operators of conditional symmetry; we
describe what kind of transformations can be applied   to  transform
 boundary conditions at infinity to those containing no conditions at infinity; and  we show that the domain geometry plays an important role in multidimensional ($n>1$) case.

Consider a BVP for the   evolution equation (\ref{2-1}) involving  conditions (\ref{2-2}) and
boundary  conditions at infinity:
\begin{equation}
\gamma_c(t,x)=\infty: \ \Gamma_c \left(t,x, u, u_x, \ldots ,
u_{x}^{({k_c})}\right) = 0, \ c = 1, 2, \ldots, p_{\infty}.\label{2-4}
\end{equation}
Here $ k_c<k $ and $p_{\infty}$ are  given numbers,  the  $\gamma_c(t,x)$ are  specified  functions
 by which the domain $(t,x)$
 on which BVP in question is defined  extends to infinity in some directions. We assume
that all  the functions arising in  (\ref{2-1})--(\ref{2-2}) and  (\ref{2-4})
and  the number of  boundary and initial conditions  are such  that  a  classical solution still exists for this
BVP.

Let us assume that the  operator
\begin{equation}
Q = \xi^0 (t,x,u)\frac{\partial}{\partial t}+\xi^i
(t,x,u)\frac{\partial}{\partial x_i} + \eta
(t,x,u)\frac{\partial}{\partial u}\label{2-5}
\end{equation}
 is a $Q$-conditional symmetry of  PDE  (\ref{2-1}), i.e.  the following  criterion is satisfied (see, e.g., \cite{bl-anco02})
\be\label{2-6}  
\mbox{\raisebox{-1.6ex}{$\stackrel{\displaystyle  
Q}{\scriptstyle k}$}}
\left(u_t - F \left(t,x, u, u_x, \ldots ,
u_{x}^{(k)}\right)\right) \Big\vert_{M}=0,
\ee
where \mbox{\raisebox{-1.6ex}{$\stackrel{\displaystyle  
Q}{\scriptstyle k}$}} is the $k$th prolongation of $Q$  and the manifold $M=\{
 u_t - F \left(t,x, u, u_x, \ldots , u_{x}^{(k)}\right)=0, \,
 Q(u)  =0\}$  with   $ Q(u) \equiv \xi^0 (t,x,u)u_t+\xi^i (t,x,u)u_{x_i} -  \eta
(t,x,u)$.

\begin{remark}Rigorously speaking,   one needs to  reduce the manifold   $M$ by adding the differential consequences of equation $Q(u)  =0$ up to order $k$, which  leads to huge technical problems in the application of the criterion obtained. However, in the case of evolution equations the resulting symmetries will be still the same provided $ \xi^0 (t,x,u) \not= 0$ in $Q$ because  each such differential consequence contains one or more  mixed derivative  of the function $u$ w.r.t.  the variables  $ t$ and $x$, while the evolution equation in question does not involve any such  mixed derivatives.
\end{remark}

Let us consider for each $c = 1, 2, \ldots, p_{\infty}$  the manifold
\be\label{2-7} \textsf{M} = \{  \gamma_c(t,x)=\infty,  \,  \Gamma_c \left(t,x, u, u_x, \ldots ,
u_{x}^{({k_c})}\right) = 0 \} \ee
in the extended space of variables $ t, x, u, u_x, \ldots ,
u_{x}^{({k_c})} $ (obviously, the space dimensionality will depend on $k_c$ and, e.g. one obtains $n+2$
 dimensional space $ (t, x, u) $ in the case of Dirichlet  boundary conditions). We assume  that there exists
  a     smooth bijective transformation of the form
\be\label{2-8}
\tau = f(t, x), \quad  y= g(t, x),   \quad w= h(t,x,u),
\ee
where $y= (y_1, \ldots, y_n) $, $f(t, x)$ and $h(t,x,u)$ are  smooth  functions and   $g(t, x)$ is a smooth vector function,
 that  maps the manifold  $\textsf{M}$ into
\be\label{2-9} \textsf{M}^* = \{  \gamma^*_c(t,x)=0,  \,  \Gamma^*_c \left(\tau, y, u, u_y, \ldots ,
u_{y}^{({k^*_{c}})}\right) = 0 \} \ee
of the same dimensionality in the  extended  space $ \tau, y, w, w_y, \ldots ,
w_{y}^{({k^*_c})} $ (here $k^*_c \leq k_c $).

\noindent {\textbf{ Definition 2.}}
BVP (\ref{2-1})--(\ref{2-2}) and  (\ref{2-4})  is $Q$-conditionally invariant under  operator (\ref{2-5})   if:
\begin{itemize}
\item[(a)]  the criterion  (\ref{2-6}) is satisfied;
\item[(b)] $Q (s_a(t,x)) = 0$ when $s_a(t,x) = 0,$ \, $B_a \vert_{ s_a(t,x) = 0}= 0,$ \, $a = 1, \ldots,p$;
\item[(c)] $
\mbox{\raisebox{-1.6ex}{$\stackrel{\displaystyle  
Q}{\scriptstyle k_a}$}} \left(B_a \left(t,x, u, u_x, \ldots , u_{x}^{(k_{a})}\right) \right) =  0 $
 when  $s_a(t,x) = 0$ and  $B_a \vert_{ s_a(t,x) = 0}= 0,$ \, $a = 1, \ldots,p$;
\item[(d)] there exists a smooth bijective transform  (\ref{2-8})  mapping $\textsf{M}$ into $\textsf{M}^*$ of the same dimensionality;
\item[(e)]  $Q^* (\gamma_c^*(\tau,y)) = 0$ when $\gamma_c^*(\tau,y) = 0$, \,$c = 1,2, \ldots,p_{\infty}$;
\item[(f)]  $
\mbox{\raisebox{-1.6ex}{$\stackrel{\displaystyle  
Q^*}{\scriptstyle k^*_c}$}} \left(\Gamma_c^*\left( \tau, y, u, u_y, \ldots ,
u_{y}^{({k^*_{c}})}\right)\right) =  0  \,  $
 when  $\gamma_c^*(\tau,y) = 0$ and  $\Gamma_c^* \vert_{ \gamma_c^*(\tau,y) = 0}= 0$, \, $c = 1,\ldots,r$,
\end{itemize}
\noindent where $\, \Gamma_c^*$ and $\gamma_c^*(\tau,y)$ are
the functions $\Gamma_c $ and
$ \frac{1}{\gamma_c(t,x)}$, respectively, expressed via the new
 variables. Moreover, the operator $Q^*$, i.e  (\ref{2-5}) in the new variables, is defined almost everywhere (i.e.  except  at  a finite number of points) on $\textsf{M}^*$.

\begin{remark}
Because any $Q$-conditional symmetry operator can be multiplied by an arbitrary function, say $s_a(t,x)$,   Definition 2 implies that  the operator $Q$ does not vanish provided $s_a(t,x) = 0$. Rigorously speaking, this restriction  is valid also for Definition 1.
\end{remark}

 This definition coincides with Definition 1 if $Q$ is a Lie symmetry operator and there are no  boundary conditions at infinity (i.e.  of the
form (\ref{2-4})). In the case of  BVPs involving boundary conditions at infinity, Definition 2 essentially generalises the definitions of Lie and conditional symmetry  proposed in \cite{ch-kov11b} and \cite{ch-2013}, respectively. In fact, those definitions are valid  only for two-dimensional  BVPs  with essentially restricted forms of boundary conditions at infinity (for example, they work for the Dirichlet conditions but  cannot be applied for the Neumann conditions as shown in Example 2 below)     because they were created using the above mentioned substitution from \cite{king91}, which is a very particular  case of  (\ref{2-8})  with
$n=1, \, \tau = t, \,  y= \frac{1}{x},   \, w= u $.


 Now we demonstrate how this definition works using simple examples.
Because each $Q$-conditional symmetry is automatically a Lie symmetry we start from an  example involving the Lie symmetry only and continue with  a second example involving pure conditional invariance.
\medskip

\textbf{Example 1.}
Consider the nonlinear  BVP modelling   heat transfer in
semi-infinite solid rod, assuming that  thermal diffusivity depends
on temperature and that  the rod is
insulated at the left endpoint. Hereafter we  neglect the initial
distribution of the temperature in the rod.
Thus
the nonlinear BVP reads as
\begin{eqnarray}
& & \frac{\partial u}{\partial t} = \frac{\partial}{\partial
x}\left(d(u)\frac{\partial u}{\partial x}\right), \ t>0, \ 0<x<+ \infty, \label{2-10} \\
& & \quad x = 0: d(u)\frac{\partial u}{\partial x} = 0, \ t > 0, \label{2-11} \\
& & \quad x = + \infty: u = u_\infty, \ t > 0, \label{2-12}
\end{eqnarray}
where $u(t,x)$ is an unknown temperature field, $d(u)$ is a thermal
diffusivity  coefficient and $u_\infty$ is a given temperature at infinity.

The maximal algebra of invariance (MAI) of the governing equation  (\ref{2-10}) is well-known and is spanned by the basic operators
 $\langle
\partial_{t},
\partial_{x}, 2t \partial_{t} + x
\partial_{x} \rangle$ provided $d(u)$ is an arbitrary function. Obviously BVP  (\ref{2-10})-- (\ref{2-12}) is  invariant w.r.t. the operator $\partial_{t}$ because the boundary conditions do not involve the time variable, while the first one affects the operator $\partial_{x}$  (see item (b) of  Definition 2).
Hence we need to examine the third operator. Items (b)-(c)  of  Definition 2 are fulfilled in the case   of the  first boundary condition, while one needs to find an appropriate  bijective transform of the form  (\ref{2-8}) to check items (d)-(f).

Let us consider the obvious change of variables
\be\label{2-13}
\tau = t, \quad  y= \frac 1x,   \quad w= u,
\ee
which maps $\textsf{M} = \{  x=\infty,  \, u = u_\infty \}$ into $\textsf{M}^* = \{  y=0,  \, w = u_\infty \}$;  both manifolds have the same dimensionality $\textsf{D}=1$  because they are
 lines in the three dimensional space of variables, i.e., item (d) is fulfilled.  Transform (\ref{2-13}) maps the operator in question to the form $2\tau \partial_{\tau} -y
\partial_{y}$  and now one easily checks that this operator satisfies  items (e)-(f)  of  Definition 2 on $\textsf{M}^*  $. Thus, BVP  (\ref{2-10})-- (\ref{2-12})  is invariant  under the  two-dimensional  MAI  $\langle
\partial_{t}, 2t \partial_{t} + x
\partial_{x}, \rangle$ provided $d(u)$ is  an arbitrary function.
Note that similarity reduction associated with the second operator of this algebra is the well-known Boltzmann one
(and could of course have been identified without use of  the definition developed here).

\medskip

\textbf{Example 2.}
Consider the reaction-diffusion-convection equation
\be\label {2-15}   \frac{\partial u}{\partial t}= \   \frac{\partial}{\partial
x}\left(u ^{m} u_x\right)+  \lambda_1
u^{m}u_{x}+ \lambda_2 u^{-m},
 \ee
 where $ \lambda_k, \ k=1,2$   and $m\neq-1, 0$ are arbitrary constants, while $u_x=\frac{\partial u}{\partial x}$.
 Let us formulate a BVP with the governing  equation (\ref{2-15})  in the domain $\Omega =\{(t,x): t>0, \ x \in
 (0, +\infty) \}$ using the Neumann  boundary conditions
  \be \label {2-16} x=0: \  u_x = \varphi(t), \ee
  and
  \be \label {2-16a} x \rightarrow +\infty : \  u_x = 0, \ee
  where $\varphi(t)$  is the specified smooth function.
  So,  (\ref{2-15})-- (\ref{2-16a}) is  a  nonlinear BVP, which is  the standard  object for investigation.
 In  \cite{ch-pl-07}, it  was proved that  (\ref{2-15})  admits the  $Q$-conditional symmetry
\be \label {2-17}  Q= \   \frac{\partial}{\partial t}
 + \lambda_2 u^{-m} \frac{\partial}{\partial u}, \ee
which is not equivalent to a  Lie symmetry provided $ \lambda_2 \not=0$.

Now we apply Definition 2 to BVP  (\ref{2-15})-- (\ref{2-16a}) in order to obtain   correctly specified constraints when this problem is conditionally invariant under  operator  (\ref{2-17}). Obviously, the first item is fulfilled by the correct choice of the operator. Item  (b) is satisfied automatically because of  the operator structure.
A non-trivial result is obtained by application of  item  (c) to  the boundary condition  (\ref{2-16}). In fact, calculating the first prolongation (i.e. $k_a=1$) of operator  (\ref{2-17})
\be \label {2-18}\mbox{\raisebox{-1.6ex}{$\stackrel{\displaystyle  
Q}{\scriptstyle 1}$}}= Q  -m \lambda_2 u^{-m-1} \frac{\partial}{\partial u_t} -  m \lambda_2 u^{-m-1} \frac{\partial}{\partial u_x}\ee
  and acting on (\ref{2-16}), one obtains the  first-order ODE
\be \label {2-19} x=0: \  \dot\varphi(t) +m \lambda_2\varphi(t)
u^{-m-1}=0,  \ee
to find the function $\varphi(t)$.
Because BVP in question   involves  the condition  at infinity (\ref{2-16a}), we  also need to examine items  (d)-- (f).
Let us consider the following  change of variables (substitution (\ref{2-13}) does not work in the case of zero Neumann conditions)
\be\label{2-20}
\tau = t, \quad  y= \frac 1x,   \quad w=  \frac u x,
\ee
which maps $\textsf{M} = \{  x=+\infty,  \, u_x = 0 \}$ into $\textsf{M}^* = \{  y=0,  \, w = 0 \}$. Since   both manifolds have the same dimensionality $\textsf{D}=1$   item (d) is fulfilled.
Transform (\ref{2-20}) maps the operator in question to the form
 \be \label {2-21}  Q^*= \   \frac{\partial}{\partial \tau}
 + \lambda_2 y^{1+m}w^{-m} \frac{\partial}{\partial w }, \ee
 and now one easily checks that this operator satisfies  items (e)-(f)  of  Definition 2 on $\textsf{M}^*  $ provided $m \in (-1,0)$. In the case $m \not\in [-1,0]$, one needs the additional constrain $y^{1+m}w^{-m} \to 0$ as $(y,w) \to (0,0)$ in order to satisfy item (f) in Definition 2 (this case is not examined here but it can be done in a similar way).

Thus, we  have shown that
 BVP  (\ref{2-15})-- (\ref{2-16a})  is $Q$-conditionally invariant under operator  (\ref{2-17}) if  and only if   condition (\ref{2-19}) and constraint  $m \in (-1,0)$  hold.

 One may note that condition  (\ref{2-19})  corresponds to a Dirichet condition and, generally speaking,
 will not be compatible with the Neumann conditions  (\ref{2-16}). Happily (but not coincidentally), there is no  contradiction in this case.
 In fact,  operator  (\ref{2-17})  generates the ansatz
 \[  u^{1+m}= \   f(x)
 + \lambda_2 (m+1)t, \]
 where $f(x)$ is an unknown function. Substituting  this ansatz into the governing equation  (\ref{2-15}) and
 solving the ordinary differential equation obtained,  one finds that  $ f(x)=C_0+C_1 e^{-\lambda_1 x}$  ($C_0$
 and $C_1$ are arbitrary constants), hence  the exact solution
  \be \label {2-22}  u= \ \left(C_0+C_1 e^{-\lambda_1 x}
 + \lambda_2 (m+1)t\right) ^{\frac{1}{1+m}}, \ee
 of  the nonlinear equation  (\ref{2-15}) is constructed. Now we need to specify the function $\varphi(t)$ using
  (\ref{2-16}), therefore  $\varphi(t)= -\frac{\lambda_1 C_1}{1+m}\left(C_0+C_1
 + \lambda_2 (m+1)t\right) ^{-\frac{m}{1+m}}$  is obtained by simple calculations. The last step is to check
 the additional condition  (\ref{2-19}), which is fulfilled identically by the function $\varphi(t)$ obtained.

Note that there is a case when the constraint (\ref{2-19})  does  not produce any boundary condition, namely  $\varphi(t)=0, \ $
i.e.  the problem  with the zero Neumann conditions (zero  flux)  on the boundary $x=0$ and at infinity  $x=+\infty$
is invariant under the  $Q$-conditional symmetry  (\ref{2-17}) provided $m \in (-1,0)$.


\section{Lie symmetry  classification of the BVPs class (\ref{0.1})--(\ref{0.3})}

Since the BVP class (\ref{0.1})--(\ref{0.3}) contains   two
arbitrary functions, $d(u)$ and  $q(t)$, the problem of Lie group
classification arises, i.e., to describe all possible Lie (or indeed
conditional) symmetries that  can  be admitted  by  BVPs from  this
class depending on the pair $(d, \,  q)$. The problem of group
classification for
 classes partial differential equations (PDEs) was  formulated by  Ovsiannikov using notions of the equivalence group $E_{eq}$ and the principal (kernel)  group of invariance  \cite{ovs}. The relevant algorithm for solving this problem, the so called  Lie--Ovsiannikov algorithm,  is well-known   (see   \cite{ovs} for details).
 During the  last decades this problem was further  studied and more efficient algorithms were worked out (see, e.g., \cite{ch-king2}, \cite{ch-king4},  \cite{vaneeva-et-al}, \cite{ch-se-ra-08} and references cited therein).
 It is widely accepted that the problem of group classification is  completely solved for the given PDE class if it has been proved that

\begin{itemize}
\item[i)] the Lie  symmetry algebras are the  maximal algebras of invariance  of   the relevant PDEs from the  list obtained;
\item[ii)] all  PDEs from the list are inequivalent with respect to a set of  transformations, which are explicitly (or implicitly) presented and, generally speaking, may not form any group;
  \item[iii)]  any other PDE from the class that   admits   a non-trivial Lie  symmetry algebra is reduced by  transformations from the set   to one of those from the list.
\end{itemize}

In \cite{ch-kov11a, ch-kov12a}  an algorithm  for  solving the group classification problem for  BVP classes was proposed. The algorithm, which is based on the concept of equivalence group of a class of BVPs, has its origins in the  Lie--Ovsiannikov algorithm.
The main steps of the  algorithm in the case  of the BVP class (\ref{0.1})--(\ref{0.3}) can be formulated as follows:
\begin{itemize}
\item[(I)] to construct  the equivalence group $E_{eq}$ of local
transformations that  transform the governing  equation (\ref{0.1})
into itself;
\item[(II)] to find the equivalence group $E_{eq}^{BVP}$
of local transformations that  transform the class of BVPs
(\ref{0.1})--(\ref{0.3}) into itself: to do this,  one extends the space of  the group $E_{eq}$
 action on the prolonged  space, where the
function $q$  arising in the boundary condition
 is treated as a new variable;
\item[(III)] to perform the group classification of
 equation (\ref{0.1}) up to local transformations generated by
the  group $E_{eq}^{BVP}$;
\item[(IV)] using Definition 2, to find the principal algebra  of invariance of the BVP  class  (\ref{0.1})--(\ref{0.3}), i.e. the algebra  admitted by each  BVP  from
this class;
\item[(V)] using Definition 2 and the results obtained in  steps (III)--(IV),
to  describe  all possible  $E_{eq}^{BVP}$-inequivalent
BVPs of the form (\ref{0.1})--(\ref{0.3})  admitting   MAIs  of higher dimensionality (depending on the pair  $(d, \,  q)$) than  the principal algebra.
\end{itemize}

The algorithm can also be applied  when one is looking for  $Q$-conditional symmetries because such symmetries cannot  generate any new group of transformations,
hence, classification can be still
carried out modulo the group $E_{eq}$.

Now we carry out the group classification of  BVPs  of the form  (\ref{0.1})--(\ref{0.3}) using
the definition and the algorithm presented above.

As the first step we find the equivalence group $E_{\mathrm{eq}}$ of the class of PDEs (\ref{0.1})
 by direct calculations and  obtain  the following result.

\begin{lemma}\label{L1}
The equivalence group $E_{\mathrm{eq}}$ of the PDEs class (\ref{0.1}) is formed by the transformations
\begin{equation*}
\tilde{t} = \alpha t + \gamma_0, \quad \tilde{x} = \beta R(\theta) x
+ \gamma, \quad \tilde{u} = \delta u + \gamma_u, \quad \tilde{d} =
\frac{\beta^2}{\alpha} \, d,
\end{equation*}
where $\alpha, \beta, \gamma_u, \gamma_i \, (i = 0, 1, 2), \delta, \theta$ are arbitrary real constants obeying
the conditions $\alpha \beta \delta \neq 0$ and $ \theta \in [- \pi, \pi)$;
 $R(\theta) = \left(\begin{array}{cc} \cos\theta & \sin\theta \\ -\sin\theta & \cos\theta \end{array} \right)$
  is the rotation matrix,   the vectors
   $\tilde{x} = \left(\begin{array}{c} \tilde{x_1} \\ \tilde{x_2} \end{array} \right)$ , \ $x = \left(\begin{array}{c} x_1 \\ x_2 \end{array} \right)$ and  $\gamma = \left(\begin{array}{c} \gamma_1 \\ \gamma_2 \end{array} \right)$.
\end{lemma}

Note that this equivalence group can be easily extracted from  paper \cite{dor-svi}, where  Lie symmetries of the class of reaction-diffusion equations
of the form
\begin{equation}\label{dor}
\frac{\partial u}{\partial t} = \nabla.\left(d(u)\nabla u \right) + Q(u),
\end{equation}
were completely described.

 In the second step, we substitute the transformations from the group $E_{\mathrm{eq}}$ into
  (\ref{0.1})--(\ref{0.3}) and  require  that those transformations  preserve the structure of the class: hence we find the set $E_{\mathrm{eq}}^{\mathrm{BVP}}$ of equivalence transformations that  are essentially different,
  using the result of Lemma 1.

\begin{lemma}\label{L2}
The equivalence group $E_{\mathrm{eq}}^{\mathrm{BVP}}$ of the class of BVPs (\ref{0.1})--(\ref{0.3}) is formed by the transformations
\begin{gather*}
\tilde{t} = \alpha t + \gamma_0, \quad \tilde{x}_1 = \beta x_1 + \gamma_1, \quad \tilde{x}_2 = \beta x_2, \quad \tilde{u} = \delta u + \gamma_u, \\ \tilde{d} = \frac{\beta^2}{\alpha} \, d, \quad \tilde{q} = \frac{\beta \delta}{\alpha} \, q,
\end{gather*}
where $\alpha>0, \gamma_u, \gamma_i \, (i = 0, 1), \delta$ and $\beta>0$  are arbitrary real constants obeying only the non-degeneracy condition  $\delta \neq 0$.
\end{lemma}

 In the third step,  we have used    the known results \cite{dor-svi} (it is interesting to  note that Lie symmetries
 of the nonlinear  equation  (\ref{0.1}) seem to have been  described  for the first time in  paper \cite{narib1970}, incidentally
  not cited so often as  \cite{dor-svi} published 13 years later) for solving  the relevant group classification problem
    in the case of  the equivalence group $E_{\mathrm{eq}}^{\mathrm{BVP}}$ and have proved the following statement.

\begin{theorem}\label{T}
All possible MAIs (up to the equivalent transformations from the group $E_{\mathrm{eq}}^{\mathrm{BVP}}$) of equation (\ref{0.1}) for any fixed
 non-negative function $d(u) \neq \mbox{const}$ are presented in Table 1. Any other equation of the form (\ref{0.1}) is reduced by an equivalence transformation from the group $E_{\mathrm{eq}}^{\mathrm{BVP}}$ to one of those given in Table 1.
\end{theorem}

 \begin{remark}
In Table 1 the following designations of the Lie symmetry operators are used:
 \be\label{0-table1}\ba{l}
 T = \partial_t, \, X_1 = \partial_{x_1}, \, X_2 = \partial_{x_2}, \, D = 2t \partial_{t} + x_a \partial_{x_a}, \\
J_{12}=x_1 \partial_{x_2}-x_2 \partial_{x_1}, \,   D_k = k t \partial_{t} - u \partial_{u}, \, D_ e =t\partial_t  - \partial_u
 \ea \ee
  while $A(x)$ and $B(x)$ (hereafter $x=(x_1, x_2)$) are an arbitrary solution of the Cauchy-Riemann system $A_{x_1} = B_{x_2}, \, A_{x_2} = - B_{x_1}$.
\end{remark}

 Now one needs to  proceed to the  final two   steps of the group classification algorithm presented above.
The result can be formulated in form of  {\it the main
theorem} (Theorem 2), which gives the complete list of the
non-equivalent BVPs  of the form  (\ref{0.1})--(\ref{0.3}) and  the
relevant MAIs.

\begin{theorem}\label{T1}
All possible MAIs (up to equivalent transformations from the group $E_{\mathrm{eq}}^{\mathrm{BVP}}$)
 of the nonlinear BVP (\ref{0.1})--(\ref{0.3}) for any fixed pair
$(d(u), q(t))$, where $d(u) \neq \mbox{const}$ are presented in Table 2.
 Any other BVP of the form (\ref{0.1})--(\ref{0.3}) is reduced by an equivalence transformation from
  the group $E_{\mathrm{eq}}^{\mathrm{BVP}}$ from Lemma 2 to one of those listed  in Table~2.
\end{theorem}

\begin{table}
\caption{Result of group classification of the class of PDEs (\ref{0.1})}
{\renewcommand{\arraystretch}{1.5}
\begin{center}
\begin{tabular}{|c|c|l|}
  \hline
  Case & $d(u)$ & Basic operators of MAI \\
  \hline \hline
  1. & $\forall$ & $AE(1,2) = \langle
  T, \, X_1, \, X_2, \,D, \, J_{12} \rangle$ \\
  2. & $u^k, k\neq 0, -1$ & $AE(1,2),
  D_ k$ \\
  3. & $u^{-1}$ & $AE(1,2),$\\
    & & $A(x) \partial_{x_1} + B(x) \partial_{x_2} - 2 A_{x_1} u \partial_{u}$ \\
    4. & $e^u$ & $AE(1,2),
  D_ e$ \\
  \hline
\end{tabular}
\end{center}}
\end{table}

\newpage

\begin{table}
\caption{Result of group classification of the class of BVPs (\ref{0.1})--(\ref{0.3})}
{\renewcommand{\arraystretch}{1.5}
\begin{center}
\begin{tabular}{|c|c|c|l|l|}
  \hline
  Case & $d(u)$ & $q(t)$ & Basic operators of MAI & Relevant  constraints \\
  \hline\hline
  1. & $\forall$ & $\forall$  & $ X_1 $ & \\
  2. & $\forall$ & $q_0 t^{-\frac{1}{2}}$  & $X_1, \, D$  & \\
  3. & $\forall$ & $q_0$  & $X_1, \, T$  & \\
  4. & $\forall$ & 0  & $X_1, \, T, \, D$  & \\
  5. & $u^k$ & $q_0 t^p$  & $X_1, \, D_{kp}$  & $k \neq -2,  \  p \neq 0$ \\
  6. & $u^k$ & $q_0 e^{\pm t}$  & $X_1, \, D_{\pm} $ & $k \neq -2$ \\
  7. & $u^k$ & $q_0$  & $X_1, \, T, \, D_{kp}$ & $p = 0$ \\
  8. & $u^k$ & 0  & $X_1, \, T, \, D, \ D_k$ & $k \neq -1$ \\
  9. & $u^{-2}$ & $\forall$  & $X_1, \, D_{\pm}$ & $k = -2$ \\
  10.& $u^{-2}$ & $q_0 t^{- \frac{1}{2}}$  & $X_1, \, D, \, D_k$ & $k = -2$ \\
  11.&  $u^{-1}$ & 0  & $X_1, \, T, \, D,
   \, X^{\infty}$ & $\mathcal{M}$\\
  12. & $e^u$ & $q_0 t^p$  & $X_1, \, D_{p}$  & $ p \neq 0$ \\
  13. & $e^u$ & $q_0 e^{\pm t}$  & $X_1, \, D_{\pm e} $ &  \\
  14. & $e^u$ & $q_0$  & $X_1, \, T, \, D_{p}$ & $p = 0$ \\
  15. & $e^u$ & 0  & $X_1, \, T, \, D, \ D_e$ &  \\
  \hline
\end{tabular}
\end{center}}
\end{table}

\begin{remark}
In Table 2 the arbitrary constant  $q_0\not=0$  and  the following designations of the Lie symmetry operators are used:
 \be\label{0-table2}\ba{l}
   D_{kp} = (k + 2) t \partial_t + [k(p+1) + 1] x_a \partial_{x_a} + (2p + 1) u \partial_u, \\ D_{\pm} =\pm (k+2) \partial_t + k  x_a \partial_{x_a} + 2 u \partial_u, \\
    D_{p} =  t \partial_t - (p-1) x_a \partial_{x_a} - (2p - 1) u \partial_u, \\ D_{\pm e} =\pm  \partial_t -  x_a \partial_{x_a} - 2  \partial_u.
 \ea \ee
In case 11, the coefficient $B(x)$ of the operator

\[ X^{\infty} = A(x) \partial_{x_1} + B(x) \partial_{x_2} - 2 A_{x_1} u \partial_{u} \]
  must satisfy  the set of conditions $\mathcal{M}$:
\begin{equation}\label{1-table2}
B(x_1, 0) =0
\end{equation}
 \begin{equation}\label{2-table2}
 x_2 \rightarrow + \infty: \, \frac{ B(x_1, x_2)}{x_2}\not=\infty,
  \quad   \frac{\partial B(x_1, x_2)}{\partial x_2}
  \not=\infty.
 \end{equation}
\end{remark}

\textbf{Proof.} The proof is based at  Definition 2, Lemma 2 and Theorem 1. According to   the algorithm described above (see steps (IV) and (V)),  we need to examine the four different cases listed in Table 1. First of  all, we  should  consider case   case 1 with the aim to find
 the principal algebra  of invariance,
i.e. the invariance  algebra, admitting by    each BVP of the form   (\ref{0.1})--(\ref{0.3}).
 Taking the most general form of the Lie symmetry in this case, one obtains
 \begin{equation}\label{3-5}
X = (\lambda_0 + 2 \lambda_3 t)\partial_t + (\lambda_1 + \lambda_3 x_1 + \lambda_4 x_2)\partial_{x_1}
+ (\lambda_2 - \lambda_4 x_1 + \lambda_3 x_2 )\partial_{x_2}
\end{equation}
where $\lambda_0, \ldots, \lambda_4$ are arbitrary real constants.
Applying  item (a)   of  Definition 2 to the first part of the  boundary condition (\ref{0.2}), we immediately obtain
\begin{equation}\label{3-6}
X(x_2)\vert_{ x_2 = 0}=0 \Leftrightarrow  \lambda_2=\lambda_4=0.
\end{equation}
To finish application of  item (a), we need the first prolongation of operator (\ref{3-5}) with $\lambda_2=\lambda_4=0$ because the boundary condition in question involves the derivative $u_{x_2}$. Hence, using the prolongation  formulae (see, e.g., \cite{olv,ovs}), we arrive at the expression
\begin{equation}\label{3-7}
\mbox{\raisebox{-1.6ex}{$\stackrel{\displaystyle  
X}{\scriptstyle 1}$}}\left. \Bigl(d(u)\frac{\partial u}{\partial x_2} - q(t)\Bigr)\right \vert_ \textsf{M}
=- \lambda_3q(t)- (\lambda_0+2\lambda_3t)
\dot q(t)=0,
\end{equation}
where $\mbox{\raisebox{-1.6ex}{$\stackrel{\displaystyle  
X}{\scriptstyle 1}$}}= X-  \lambda_3(2u_t\partial_{u_t}+u_{x_a} \partial_{u_{x_a}})  $ and
\be\label{3-8} \textsf{M} = \{   x_2 = 0, \,  d(u)\frac{\partial u}{\partial x_2} = q(t) \}. \ee
Obviously, the zero flux case $q(t)=0$ does not produce any constraints, hence $\lambda_0, \lambda_1$ and $\lambda_3$ are arbitrary, i.e., the relevant BVP is invariant under a 3-dimensional MAI (see case 4 of Table 4).
Rather simple analysis of  the linear ODE  $\lambda_3q(t)+ (\lambda_0+2\lambda_3t)
\dot q(t)=0 $ with non-zero  $q(t)$ immediately leads   to three different possibilities only:
\begin{itemize}
\item[i)] if $q(t)$ is an arbitrary function then  $\lambda_0=\lambda_3=0$, i.e., $X=X_1$;
 \item[ii)] if $q(t)=q_0/\sqrt{t+\lambda_0^*}$  with   $\lambda_0^*=\lambda_0 / (2\lambda_3)$   then   $X= \lambda_0 T + \lambda_1 X_1+\lambda_3 D$ (here $\lambda_0$  and $\lambda_3 \not=0$ are no longer arbitrary);
      \item[iii)] if $q(t)=q_0$,    $q_0$ being a constant, then  $\lambda_3=0$, i.e., $X= \lambda_0 \partial_t +\lambda_2\partial_{x_2}$.
     \end{itemize}
   The function $q(t)$ and the operator $X$ arising in   item (ii) can be simplified using Lemma 2 (see transformation for $t$) as follows
   $q(t) \rightarrowtail q_0/\sqrt{t}, \,  X  \rightarrowtail \lambda_1 X_1+  D$.
   Now we need to find an appropriate transform of the form (\ref{2-8}).
      Let us consider the transformation
 \be\label{3-1}
\tau = t, \quad  y_1=x_1,   \quad y_2=x_2^{-\epsilon},  \epsilon>0, \quad u= x_2 w,
\ee
which transforms the manifold
\be\label{3-2} \textsf{M} = \{ x_2 \rightarrow + \infty, \   \frac{\partial u}{\partial x_2}= 0  \} \ee
into
\be\label{3-3} \textsf{M}^* = \{ y_2 =0, \   w= 0  \}, \ee
provided  the function $w$  is differentiable at $ y_2 =0$ (it should be noted that  transforming
only
$y$ (keeping the other variables  the same) does not work in that sense).

Now one realizes that  items (e)--(f)   of  Definition 2  are  automatically fulfilled if  $q(t)$ is an arbitrary function and $X=X_1$ (see  item (i) above), hence we have found
 the principal algebra  of invariance of the BVPs  class  (\ref{0.1})--(\ref{0.3}) and one is listed in the case 1 of Table.
 To examine the other   two items one needs to express the relevant operators via the new variables. In particular, the operator $X=\lambda_1 X_1+  D$  takes the form
\be\label{3-9}  X^* = 2t \partial_{t} + ( \lambda_1 +x_1) \partial_{x_1} -  \epsilon y_2 \partial_{y_2}- w\partial_{w}.
\ee
Obviously items (e)--(f)   of  Definition 2  are   automatically fulfilled for operator  (\ref{3-9}) provided the boundary condition is
 giving $ \textsf{M}^*$, hence  the case 2 of Table 2 is derived.  Obviously, the third possibility for the function $q(t)=q_0$ leads to  case 3 of Table 2.
Thus, case 1 of Table 1 produces the principal algebra of invariance of the BVPs  class  (\ref{0.1})--(\ref{0.3}) and three  extensions depending on the function $q(t)$ (see cases 1--4 in Table 2).

Case 2  of  Table 1 can be examined in quite a similar way as we have done above for case 1.
Application of Definition 2 and  Lemma 2 leads to the 6 different cases listed in Table 2
(see  $5, \dots, 10$).
 It should be  stressed that the power $k=-2$ is   a special one (but   not  one for    Lie
invariance of  the governing PDE!)  and leads to two additional
cases 9-10 (see an  analogous result
 for a (1+1)-dimensional BVP in \cite{kov2011,ch-kov12a}).

The most complicated examination   is in case 3 of Table 1 because one needs to  analyse an infinite-dimensional
Lie algebra.
In this case,
 MAI of equation (\ref{0.1}) is spanned by the following operators
\begin{gather*}
T = \partial_t, \, X_a = \partial_{x_a}, \, J = x_2 \partial_{x_1} - x_1 \partial_{x_2}, \, D = 2 t \partial_t + x_a \partial_{x_a}, \, D_{-1} = t \partial_t + u \partial_u,\\ X^{\infty} = A(x_1,x_2) \partial_{x_1} + B(x_1,x_2) \partial_{x_2} - 2 A_{x_1} u \partial_{u}.
\end{gather*}
Hence, the most general form of Lie symmetry operator is
\be\label{3-10}\ba{l}
 X =
(\lambda_0 + (2 \lambda_4 + \lambda_5)t) \partial_t + (\lambda_1 + \lambda_4 x_1 + \lambda_3 x_2 + \lambda_6 A(x_1,x_2)) \partial_{x_1} \\
+ (\lambda_2 - \lambda_3 x_1 + \lambda_4 x_2 + \lambda_6 B(x_1,x_2)) \partial_{x_2} - (2 \lambda_6 A_{x_1} - \lambda_5) u \partial_u,
\ea \ee
where $\lambda_0, \ldots, \lambda_6$ are arbitrary real constant.

 Obviously, we should assume $\lambda_6 \neq 0$, otherwise  particular cases  of  the results already derived
 for $d(u)=u^k$ will be obtained.
  First of all, we simplify operator (\ref{3-10}) as follows. Because the functions $A(x_1,x_2)$ and $B(x_1,x_2)$
  are  harmonic, then  we can construct  the  functions
  $\bar{A} = \lambda_1 + \lambda_3 x_2 + \lambda_6 A(x_1,x_2)$ and $\bar{B} = \lambda_2 - \lambda_3 x_1  + \lambda_6 B(x_1,x_2)$. It can be easily  seen that  the functions $\bar{A}$ and $\bar{B}$  are also harmonic. Thus, without losing generality,
 the operator $X$ reduces to the form
\begin{equation}\label{3-11}
X=
(\lambda_0 +(2 \lambda_4 + \lambda_5)  t) \partial_t + (\bar{A}(x_1,x_2)+ \lambda_4 x_1) \partial_{x_1} + (\bar{B}(x_1,x_2)+ \lambda_4 x_2) \partial_{x_2} - (2 \bar{A}_{x_1} - \lambda_5) u \partial_u.
\end{equation}

Applying  items (b)--(c)   of  Definition 2 to the  boundary condition (\ref{0.2}), we  obtain
\begin{equation}\label{3-12}
\left.  X(x_2) \right \vert_{x_2 = 0} = 0, \quad \mbox{\raisebox{-1.6ex}{$\stackrel{\displaystyle  
X}{\scriptstyle 1}$}}  \left.  \left(u^{-1} \frac{\partial u}{\partial x_2} - q(t)\right) \right \vert_{\textsf{M}} = 0,
\end{equation}
where  the first prolongation of the operator $X$  has the form $ \mbox{\raisebox{-1.6ex}{$\stackrel{\displaystyle  
X}{\scriptstyle 1}$}}= X + \rho_0\partial_{u_t}+\rho_{a} \partial_{u_{x_a}}$
and $\textsf{M}$ is defined in (\ref{3-8}). We need to calculate only the coefficient $\rho_{2}$
because  (\ref{3-12}) does not  involve  any other  derivatives.
Since the known formulae mentioned above produce in the case of  operator (\ref{3-11}):
  \begin{equation}\label{3-13}
  \rho_{2}=-2 \frac{\partial^2 \bar{B}}{\partial x_2^2}u +\Bigl(\lambda_4 + \lambda_5
  - 3\frac{\partial \bar{B}}{\partial x_2}\Bigr)u_{x_2}- \frac{\partial \bar{A}}{\partial x_2}u_{x_1}
  \end{equation}
   the invariance conditions  (\ref{3-12})  are simplified to the form
 \begin{equation}\label{3-14}
  \bar{B}(x_1, 0) = 0, \quad \frac{\partial \bar{A}(x_1, 0)}{\partial x_2} = 0, \quad  \frac{\partial^2 \bar{B}(x_1, 0)}{\partial x_2^2}=0
\end{equation}
provided $q(t) = 0$
 (we  remind the reader that the functions $\bar{A}$ and $\bar{B}$  satisfy  the Cauchy-Riemann system).
 Obviously, conditions (\ref{3-14}) are equivalent to  this
(\ref{1-table2}) if one takes into account that the second and third
equations in  (\ref{3-14}) are direct consequences of the first
equation.

To finish examination of  case 3 of Table 1 when the nonlinear  BVP in question  involves  zero flux  $q(t) = 0$, one needs to check invariance of the boundary condition (\ref{0.3}). Hence, using again  transformation (\ref{3-1}) and applying items   (f) and (e)  of  Definition 2,
one arrives at the restrictions
\begin{equation}\label{3-15}
y_2=0: \, \bar{B}(x_1, y_2^{-1/\epsilon}) y_2^{1+ 1/\epsilon}= 0
\end{equation}
and
\begin{equation}\label{3-16}
\left. \Bigl( \Bigl(\bar{B}(x_1, y_2^{-1/\epsilon}) y_2^{1/\epsilon} +2\frac{\partial \bar{A}(x_1, y_2^{-1/\epsilon})} {\partial{x_1}}\Bigr)w\Bigr)
\right \vert_{ \textsf{M}^*} =0.
\end{equation}
Because transformation (\ref{3-1}) is bijective and differentiable, formulae (\ref{3-15})--(\ref{3-16}) are equivalent to
(\ref{2-table2}).  Hence, we have  proved that BVP   (\ref{0.1})--(\ref{0.3}) with $d(u)=u^{-1}$ and  $q(t) = 0$  admits the operator
(\ref{3-11}) provided restrictions (\ref{1-table2})--(\ref{2-table2}) are fulfilled. This immediately leads to the result presented in case 11 of Table 2  (the rotation operator $J_{12}$ must  be excluded because its coefficient $B =-x_1$ does not satisfy (\ref{1-table2})).

The invariance  conditions (\ref{3-12}) lead to more complicated analysis if $q(t) \not= 0$.  We omit here the relevant routine analysis  and present the result only: the restrictions obtained on the functions $\bar{A}$ and $\bar{B}$  lead to the correctly specified functions $q(t)$  and
MAIs  listed in cases 5--7  with $k=-1$  of Table 2   only.
Thus, examination of  case 3 of Table 1 when  $d(u) = u^{-1}$ is completed.

Finally, case 4  of  Table 1 should  be analysed. It turns out that the  results obtained are  very similar to  case 2  of  Table 1 when $k \not=-2$,  therefore  MAIs of the same dimensionality  and  the  fluxes  $q(t) $  of the same forms were derived (see cases 12--15 in Table 2).

The proof is now completed. \hfill $\blacksquare$

\bigskip

 While restrictions (\ref{1-table2})--(\ref{2-table2}) on the harmonic functions $A$ and $B$  are very strong,
  MAI of  the problem  in case 11 is still infinite-dimensional. Since the real and imaginary
   parts of the complex function $z^{-n}$
with {\it arbitrary } $n=1,2,3,...$ generates the operator of the form $X^\infty$,
 which is a symmetry of  BVP (\ref{0.1})--(\ref{0.3}) with $d(u)=u^{-1}$ and  $q(t)=0$.
  Note  that here we allow singular behaviour of $X^\infty$ at the  origin $(x_1,x_2)=(0,0).$

 \noindent {\textbf{ Example 3.}}
The complex function $z^{-1}$ generates the operator
 \begin{equation}\label{3-0} \frac{x_1}{x_1^2+x_2^2} \partial_{x_1} - \frac{x_2}{x_1^2+x_2^2}  \partial_{x_2} + 2 \frac{x_1^2-x_2^2}{(x_1^2+x_2^2)^2} u \partial_{u}   \end{equation}
Applying  items (b) and (c)  of  Definition 2, one obtains  (\ref{3-12})  and   (\ref{3-13})  with $\lambda_4 = \lambda_5=0$ and
$ \bar{A}=\frac{x_1}{x_1^2+x_2^2}, \, \bar{B}=-\frac{x_2}{x_1^2+x_2^2}$. Now one easily checks that the invariance conditions are satisfied  (\ref{3-12}) because the given functions $ \bar{A}$ and $ \bar{B}$ fulfil conditions  (\ref{3-14}).

 Let us consider the transformation  (\ref{3-1}). As   indicated above, one
 transforms the manifold  (\ref{3-2})  into  (\ref{3-3}).
Simultaneously  operator (\ref{3-0}) takes the form
 \begin{equation}\label{3-4} \frac{y_1}{|y|^2_\epsilon} \partial_{y_1} + \frac{1}{|y|^2_\epsilon} (\epsilon y_2\partial_{y_2} + w \partial_{w}) +2 \frac{y_1^2-y_2^{-2/\epsilon}}{|y|^4_\epsilon} w \partial_{w},   \end{equation}
where $|y|^2_\epsilon =y_1^2+y_2^{-2/\epsilon}$.

After directly   checking   items (d)--(f)  of  Definition 2  in the case of manifold (\ref{3-3}) and operator (\ref{3-4}),  one concludes that (\ref{3-0}) is a Lie symmetry  operator of BVP (\ref{0.1})--(\ref{0.3}) with $d(u)=u^{-1}$ and  $q(t)=0$.

\medskip

 We  conclude this section  by presenting the following observation. Let as replace the last condition in BVP (\ref{0.1})--(\ref{0.3}) by
\begin{equation}\label{3-20}
 x_2 \rightarrow + \infty:  d(u)\frac{\partial u}{\partial x_2}= 0 \ee
 which is more   usually
 adopted  in  applications.
It can be easily checked by direct calculations (each Lie symmetry operator generates the corresponding Lie group of transformations) that the results presented in Table 2 are still valid for BVPs of the form (\ref{0.1}), (\ref{0.2})  and (\ref{3-20}). Moreover, the  assumption  $d(u)\not= 0$  for  $x_2 \rightarrow + \infty$ is not important (though it is of course significant with respect to BVP theory).  However, one should
  ideally show that there are no cases other than those presented in Table 2. Unfortunately, this  is a non-trivial task, in particular, transformation    (\ref{3-1})
does not  work in all  cases as above.  For example, to examine case of the power-law diffusivity $d(u)= u^k, \  k\not=-1$ one could use the  transformation
 \[
\tau = t, \quad  y_1=x_1,   \quad y_2=x_2^{-\epsilon},  \epsilon>0, \quad u= x^{\frac{1}{k+1}}_2 w,
\]
which transforms $\textsf{M}$  into
\[ \textsf{M}^* = \{ y_2 =0, \   w^{k+1}= 0  \}. \]


\section{Lie symmetry  reduction of some  BVPs  of the form  (\ref{0.1})--(\ref{0.3})}

First of all,  it should be noted that each BVP  of the form  (\ref{0.1})--(\ref{0.3}) reduces to
 a (1+1)-dimensional problem using the operator $X_1=\partial_{x_1}$. However, the problem obtained is
  simply the corresponding (1+1)-dimensional one, with no dependance on  $x_1$;  hence we do not consider
 such a reduction below.

Another special case arises for each  BVP    (\ref{0.1})--(\ref{0.3}) with $q(t)=q_0$ (case 3 of Table 2)
when the problem reduces to the stationary one using  the operator $T=\partial_{t}$:
\begin{eqnarray}
& &  \nabla.\left(d(U)\nabla U \right)=0, \ \  \label{0.1*} \\
& & \quad x_2 = 0: d(U)\frac{\partial U}{\partial x_2} = q_0, \label{0.2*} \\
& & \quad x_2 \rightarrow + \infty:  \frac{\partial U}{\partial x_2}= 0 , \label{0.3*}
\end{eqnarray}
where $U(x_1,x_2)$ is an unknown function.  BVP    (\ref{0.1*})--(\ref{0.3*}) is linearisable via the Kirchhoff substitution $W=\int d(U)dU$ and the linear problem obtained can be treated by the classical methods for solving linear problems for the Laplace equation.

A brief analysis of Table 2  shows that seven cases when the
relevant problems are invariant under MAI of dimensionality three
and higher are the most interesting because a few different
reductions to BVPs of lower dimensionality can be obtained.
Obviously the most complicated case occurs for the critical exponent
$k=-1$ (see case 11)  and we are going to
 treat in detail this one
elsewhere. On the other hand, cases 7 and 8 seem to be the most
interesting because the power diffusivity $u^k$ is very common in
applications and describe a wide range of phenomena depending on the
value of $k$.

Let us consider case 7. Because the operator $D_{kp}$ with $p=0$ has the form

\begin{equation}\label{4-0}
 D_{k0} = (k + 2) t \partial_t + (k+ 1) x_a \partial_{x_a} + u \partial_u \ee
 one needs to consider three different cases
 \begin{itemize}
\item[i)] $k\not=-1, \ -2;$
\item[ii)] $k= -2;$
  \item[iii)] $k=-1.$
\end{itemize}

In the first case,  the Lie algebra $\langle  X_1, \, T, \, D_{k0}
\rangle$ leads only  to two essentially different reductions, via
the operators  $ T+ vX_1, \ v \in \mathbb{R} $  and $D_{k0}$.
Obviously the operator $ T+ vX_1, \ v \in \mathbb{R} $ generates the
travelling-wave ansatz
\begin{equation}\label{4-0*}
u= \ \phi(y,x_2), \quad  y= x_1-vt
\ee
which reduces the nonlinear BVP
\begin{eqnarray}
& & \frac{\partial u}{\partial t} = \nabla.\left(u^k\nabla u \right), \ \ (x_1,x_2) \in \Omega, \, t \in \mathbb{R}, \label{4-1} \\
& & \quad x_2 = 0:  u^k\frac{\partial u}{\partial x_2} = q_0, \label{4-2} \\
& & \quad x_2 \rightarrow + \infty:  \frac{\partial u}{\partial x_2}= 0 , \label{4-3}
\end{eqnarray}
to the (1+1)-dimensional elliptic problem
\begin{eqnarray}
& &  \label{4-4} -v\phi_{ y}= (\phi^k\phi_{ y})_y + (\phi^k\phi_{ x_2})_{x_2} \\
& & \quad x_2 = 0:  \phi^k\phi_{ x_2} = q_0, \label{4-5} \\
& & \quad x_2 \rightarrow + \infty:  \phi_{ x_2}= 0, \label{4-6}
\end{eqnarray}
where $ \phi$ is an unknown function (hereafter
 subscripts on $ \phi$ denote differentiation w.r.t. the relevant variables).
 Like other cases described below, the relevance of the function
$\phi$ to a specific BVP will depend on the behaviour at infinity of
the initial data and we shall make no attempt to explore such
matters here in detail.

The operator $D_{k0}$ generates a more complicated ansatz
\begin{equation}\label{4-7}
u= \ t^{\frac{1}{k+2}}\phi(\omega_1, \omega_2), \quad   \omega_a=
x_a t^{-\gamma}, \  \gamma=\frac{k+1}{k+2}. \ee
 After substituting
ansatz (\ref{4-7}) into BVP (\ref{4-1})--(\ref{4-3}), direct
calculations show that one obtains the (1+1)-dimensional elliptic
problem
\begin{eqnarray}
& &  \label{4-8} \frac{1}{k+2} \phi - \gamma \omega_a\phi_{ \omega_a}=  (\phi^k\phi_{ \omega_a})_{\omega_a} \\
& & \quad \omega_2 = 0:  \phi^k\phi_{ \omega_2} = q_0, \label{4-9} \\
& & \quad  \omega_2 \rightarrow + \infty:  \phi_{ \omega_2}= 0 \label{4-10}
\end{eqnarray}
(hereafter  summation  is assumed over the repeated index $a =1,2$).

In  case {\it (ii)},  the Lie algebra $\langle  X_1, \, T, \, - x_a \partial_{x_a}+u \partial_u \rangle$ leads   to three  essential different reductions, via the operators $ T+ vX_1, \  v \in \mathbb{R} $,  \ $\frac{1}{\lambda}T- x_a \partial_{x_a}+u \partial_u , \ \lambda \not=0$ and $ - x_a \partial_{x_a}+u \partial_u$. Obviously the operator $ T+ vX_1, \ v \in \mathbb{R} $ leads to the same ansatz as in case {\it (i)}, hence  BVP (\ref{4-4})--(\ref{4-6}) with $k=-2$ is obtained.

The operator  $\frac{1}{\lambda}T- x_a \partial_{x_a}+u \partial_u$  generates a new  ansatz of the form
\begin{equation}\label{4-11}
u= \ e^{\lambda t}\phi(\omega_1, \omega_2), \quad   \omega_a= x_a
e^{\lambda t}, \  \lambda \not=0, \ee
 which reduces BVP
(\ref{4-1})--(\ref{4-3}) with $k=-2$ to the (1+1)-dimensional
problem
\begin{eqnarray}
& &  \label{4-12}  \lambda \phi  + \lambda \omega_a\phi_{ \omega_a}=  (\phi^{-2}\phi_{ \omega_a})_{\omega_a} \\
& & \quad \omega_2 = 0:  \phi^{-2}\phi_{ \omega_2} = q_0, \label{4-13} \\
& & \quad \omega_2 \rightarrow + \infty:  \phi_{ \omega_2}= 0. \label{4-14}
\end{eqnarray}

As is well-known, the singular nature of the diffusivity in  (\ref{4-12}) as $\phi \rightarrow 0$  prevents immediate  physical interpretation of such reductions, but it is
 worth noting that the reduction   (\ref{4-11}) applies for a continuum of values of the similarity exponent $\lambda$.
The one-dimensional case is instructive here, yielding  the equation
\begin{equation}\label{4-12a}  q_0 + \lambda \omega_2\phi=  \phi^{-2}\phi_{ \omega_2}. \ee
This ODE is easely solved by setting $\psi =\omega_2 \phi$  and  its general solution
can be presented in the implicit form (for $q_0=0$  and  $\lambda=\frac{q_0^2}{ 4}$  the solutions  are obvious)
\begin{equation}\label{4-12b}
C=\frac{\phi}{\sqrt{\lambda (\omega_2\phi)^2 +q_0 \omega_2\phi
+1}}\left(\frac{q_\lambda +q_0+  2\lambda \omega_2\phi} {q_\lambda
-q_0-  2\lambda \omega_2\phi} \right)^{\frac{q_0}{2q_\lambda}}, \ee
where $C$ is an arbitrary non-zero constant and
$q_\lambda \equiv \sqrt{q_0^2-4\lambda}$. Note that  we need
$\lambda<q_0^2/4$ in order to obtain a real solution. Because
solution (\ref{4-12b}) should  satisfy also the condition at
infinity  (\ref{4-14}) we need to analyse it  as $\omega_2
\rightarrow + \infty$. Indeed, it can be noted that
\begin{equation}\label{4-12c}
\phi \sim \frac{\phi_\infty}{\omega_2}, \quad \omega_2 \rightarrow + \infty, \ee
where $\phi_\infty$ is a solution of the quadratic equation
\begin{equation}\label{4-12d}\lambda \phi_\infty^2+q_0 \phi_\infty+1=0 \ee
(there are two roots and which   should be used depends on sign of $q_0$).
Thus, we conclude that
\begin{equation}\label{4-12e}
u \sim \frac{\phi_\infty}{x_2}, \quad x_2 \rightarrow + \infty,  t=0\ee
whereby  the similarity exponent  $\lambda \leq q_0^2/4$ is  determined  in terms of this initial data via  (\ref{4-12d}).

The operator  $- x_a \partial_{x_a}+u \partial_u$  generates the   ansatz
\begin{equation}\label{4-15}
u= \ x_1^{-1}\phi(t, z), \quad   z=\frac{x_2}{ x_1}, \ee reducing
the (1+2)-dimensional BVP in question to  the (1+1)-dimensional
parabolic problem
\begin{eqnarray}
& &  \label{4-16} \phi_{ t}=  (\phi^{-2}\phi_{ z})_{z} + z \Bigl(\phi^{-1} +  z\phi^{-2}\phi_{ z} \Bigr)_z  \\
& & \quad z = 0:  \phi^{-2}\phi_{ z} = q_0, \label{4-17} \\
& & \quad  z \rightarrow + \infty:  \phi_{ z}= 0. \label{4-18}
\end{eqnarray}

\begin{remark}
The reduced BVP (\ref{4-16})--(\ref{4-18})  was derived under assumption $x_1>0$.
In the case $x_1<0$, the same problem is obtained  but $z \rightarrow + \infty$ should be replaced by
 $z \rightarrow - \infty$.
\end{remark}

Finally, we examine  case {\it (iii)},  in which the Lie algebra $\langle  X_1, \, T, \, t \partial_{t}+u \partial_u \rangle$ arises.
There are only  two  essentially different reductions, via the operators
 $ T+ vX_1 $ and  $ \lambda X_1+ t \partial_{t}+u \partial_u$ where $ (v,\lambda) \in \mathbb{R}^2$.
  The first one again leads to a particular case of BVP (\ref{4-4})--(\ref{4-6}) with $k=-1$, while
  the second generates a new  (1+1)-dimensional elliptic  problem of the form
\begin{eqnarray}
& &  \label{4-19}  \phi - \lambda \phi_{ w}=  (\phi^{-1}\phi_{ w})_{w}+ (\phi^{-1}\phi_{ x_2})_{x_2} \\
& & \quad x_2 = 0:  \phi^{-1}\phi_{ x_2} = q_0, \label{4-20} \\
& & \quad x_2 \rightarrow + \infty:  \phi_{ x_2}= 0, \label{4-21}
\end{eqnarray}
where
\begin{equation}\label{4-22}
u= \ t\phi(w,x_2), \quad   w= x_1 - \lambda \log t.
\ee
It should be noted that BVP  (\ref{4-19})--(\ref{4-21}) with $\lambda=0$  is equivalent to the problem (provided $\phi=e^\psi \geq 0$)
\begin{eqnarray}
&   \label{4-23}  e^\psi=  \Delta \psi \\
&   x_2 = 0:  \psi _{ x_2} = q_0, \label{4-24} \\
&   x_2 \rightarrow + \infty:  \psi_{ x_2}= 0, \label{4-25}
\end{eqnarray}
where (\ref{4-23})  is known Liouville's equation,  which has been widely  studied
   for many years (see, e.g. the  books \cite{du-no-fo1992, henrici1993})  and its general solution (for $n=2$)
    is well-known.

 Now we
  make the following observation: while BVP in question is a  parabolic problem, all the (1+1)-dimensional BVPs obtained  (excepting  (\ref{4-16})--(\ref{4-18}))  are elliptic.
Each of the (1+1)-dimensional BVPs derived above can be further analysed  by
symmetry based, asymptotical   and numerical methods,  and  we shall investigate such matters  in a forthcoming  paper. Here we  present
an interesting example only.

\textbf{Example 4.} 
It is appropriate  to touch
 on the large-time behaviour of BVP (\ref{4-16})-(\ref{4-18}). This is best done in polar coordinate. From the symmetry point of view,  it means that one uses the ansatz
\begin{equation}\label{4-15*}
u= \ r^{-1}v(t,  \theta), \quad   \theta=\arctan\frac{x_2}{ x_1}, \, r^{2}=  x_1^2+ x_2^2, \ee
which is equivalent to (\ref{4-15}). As result the  reduced BVP takes form

\begin{eqnarray}
& &  \label{4-16*} v_{ t}=  (v^{-2}v_{  \theta})_{ \theta} -v^{-1}  \\
& & \quad z = 0:  v^{-2}v_{ \theta } = q_0, \label{4-17*} \\
& & \quad  \theta \rightarrow \frac{\pi^-}{ 2} :  \frac{\cos \theta v_{\theta}}{ v} \rightarrow 1. \label{4-18*}
\end{eqnarray}
Note that  the  conditions  $ \theta \rightarrow \frac{\pi^-}{ 2} : v = 0, \ \cos \theta v_{ \theta }\rightarrow 0$, which formally can be also used instead of (\ref{4-18*}),  are  inappropriate  because (\ref{4-16*}) has no smooth solutions satisfying $v=0$  at any finite $\theta$.

The boundary condition (\ref{4-18*}) implies that   $ v \rightarrow +\infty$ as $ \theta \rightarrow \frac{\pi^-}{ 2}$ (this allows $d(u)\equiv u^{-2} = 0$ when $x_2 \rightarrow + \infty$); this and   (\ref{4-17*}) immediately lead to  the conservation law
\begin{equation}\label{4-27}
\frac{d}{ dt}\int^{\frac{\pi}{ 2}}_0 v \cos\theta d\theta =-q_0,
\ee
 where  $v(t,\theta) \cos\theta = \phi(t,\tan\theta)$ (see (\ref{4-15})).
Then the following three-layer structure is a plausible description of the behaviour of $v$ as $t \rightarrow + \infty$.

{\it (A)} Outer region $\theta=O(1)$. Here $v \sim  \sqrt{ t}\Psi(\theta)$, which reduces  (\ref{4-16*}) to
\[ \frac{1}{ 2}\Psi=  (\Psi^{-2}\Psi_{ \theta})_{\theta}-\Psi^{-1},\]
hence one obtains
\[ \Psi  \sim \frac{1}{\theta\sqrt{-\log\theta} }  \quad  as  \quad \theta \rightarrow 0^+,  \quad
\Psi  \sim \frac{1}{(\frac{\pi}{ 2}-\theta)\sqrt{-\log(\frac{\pi}{ 2}-\theta)} }   \quad as  \quad \theta \rightarrow \frac{\pi^-}{ 2}.  \]

{\it (B)} Transition region $\Theta=O(1)$, where $-\lambda< \Theta<0$. Here we set
$v=\theta^{-1}\sigma(t,\zeta), \, \zeta=\log\theta$ in order  to obtain from  (\ref{4-16*}) the equation
(note the  term $v^{-1}= \theta \sigma ^{-1} $  is negligible under these scalings)
\begin{equation} \label{4-28} \sigma_{ t}=  (\sigma^{-2}\sigma_{ \zeta})_{ \zeta} +
\sigma^{-2} \sigma_{\zeta }. \ee

Whereby the middle term in (\ref{4-28}) is negligible (for large time)  and $\sigma \sim \Sigma(\Theta), \, \Theta=\zeta t^{-1}$, one obtains the equation
\[-\Theta\Sigma_{ \Theta}=\Sigma^{-2}\Sigma_{ \Theta}\]
with the solution $\Sigma=\frac{1}{\sqrt{-\Theta}}.$
The region dominates the integral in (\ref{4-27}), whereby
\[ \int^0_{-\lambda}\frac{ d\Theta}{\sqrt{-\Theta}} = -q_0 \]
so that $\lambda$ (which plays a crucial role in the inner region) is given by $\lambda =\frac{q_0^2}{ 4}$, a conclusion that also follows by other arguments.

{\it (C)} Inner region
$y\equiv e^{\lambda t}\theta=O(1)$.
 Here we  introduce the variable
\[ v \sim e^{\lambda t}V(y) \]
(the large-time solution behaviour may in fact also involve algebraic dependence on $t$, such refinements are of little importance here). Using these variables and neglecting the final term in  (\ref{4-16*}), one obtains the equation
\[ (V^{-2}V_{  y})_{ y}= \lambda V +  \lambda yV_y, \]
which is equivalent to the first order ODE
\begin{equation}\label{4-16**}  V^{-2}V_{  y}= \lambda yV +q_0,
\ee
if one takes into account (\ref{4-17*}). Now we note that ODE (\ref{4-16**}) coincides with  (\ref{4-12a}),
which was analysed above. In particular, the exponent $\lambda =\frac{q_0^2}{ 4}$ is associated to a repeated-root condition (see  (\ref{4-12d})). Thus, we obtain
\begin{equation} \label{4-29}
V \sim \frac{2}{-q_0y}, \quad as \quad  y \rightarrow + \infty,\ee
which matches with the region {\it (B)} and provides an alternative route to the derivation of the value of the similarity exponent $\lambda$.


\section{Conditional  symmetry  classification of the BVPs class (\ref{0.1})--(\ref{0.3})}

$Q$-conditional (nonclassical)  symmetries  of  the class of  (1+2)-dimensional  heat equations  (\ref{0.1})
were described in paper \cite{arr-goard-br}.
In contrast  to the  (1+1)-dimensional case, the result  is  very simple: in the case of  the $Q$-conditional
 symmetry operator   (\ref{2-3})  with $\xi^0 (t,x)\not=0$, there is only a unique nonlinear  equation from this
  class admitting a conditional symmetry. Any other nonlinear heat equation admits conditional symmetry operators
   of the form (\ref{2-3}), which are equivalent to the relevant Lie symmetry operators.
   In the case of  $Q$-conditional symmetry operator   (\ref{2-3})  with $\xi^0 (t,x)=0$,
   the  system of determining equations is analysed in \cite{arr-goard-br}(see the system (3.30) therein) and
    their conclusion  is as follows: each known solution   of the system leads again
     to a Lie symmetry and they were not  able to construct any other solution.

Let us consider   the equation
\begin{equation}\label{5-0}
\frac{\partial u}{\partial t} = \nabla.\left(u^{-1/2}\nabla u \right)  \end{equation}
 ant its  conditional symmetry
 \begin{equation}\label{5-1}  Q= \frac{\partial }{\partial t} +
 2h(x_1,x_2)\sqrt u \frac{\partial }{\partial u},
 \end{equation}
where the function $h$ is an arbitrary solution of the nonlinear equation $\Delta h=h^2$ (in \cite{arr-goard-br}
these formulae have a slightly different  form because in the very beginning  the authors applied  the Kirchhoff
 transformation to  (\ref{0.1})).
Now we  examine BVP  (\ref{5-0}), (\ref{0.2}) with $d(u)=u^{-1/2}$  and (\ref{0.3}) using Definition 2.
 Obviously items (a) and (b) are automatically  fulfilled. To fulfill item (c) one needs the first prolongation
  of operator   (\ref{5-1})
 \begin{equation}\label{5-2}  \mbox{\raisebox{-1.1ex}{$\stackrel{\displaystyle  
Q}{\scriptstyle 1}$}} = \frac{\partial }{\partial t} +{2}h \sqrt u \frac{\partial }{\partial u} + \Bigl(2h_{x_2}\sqrt u  +h u^{-1/2} u_{x_2}\Bigr)\frac{\partial}{\partial {u_{x_2}}}.
 \end{equation}
Applying this operator to the boundary condition  (\ref{0.2}) with $d(u)=u^{-1/2}$, we arrive at the  equation
 \begin{equation}\label{5-3}
 x_2=0: \, q '(t)= {2}h_{x_2},
  \end{equation}
  what immediately gives
   \begin{equation}\label{5-4}
    q(t)= q_0+{2} q_1 t, \quad  h_{x_2}(x_1,0)=q_1  \end{equation}
    where $q_0$ and $q_1 $ are arbitrary constants.
    Finally, we can use again transformation (\ref{3-1})  for examination of items (d)-(f) and direct checking shows that a sufficient condition is that
  the function $h$ be bounded as $x_2 \rightarrow \infty$.

    Thus, BVP  (\ref{5-0}), (\ref{0.2}) with $d(u)=u^{-1/2}$  and (\ref{0.3}) is $Q$-conditionally invariant
    only in the case of linear flux $q(t)$ (see (\ref{5-4})) and the relevant conditional symmetry
     operator possesses the form (\ref{5-1}) where the
    function $h$  solves the initial problem
     \begin{equation}\label{5-5}
     \Delta h=h^2 , \quad  h_{x_2}(x_1,0)=q_1.
     \end{equation}

\begin{remark}Because each conditional symmetry operator (\ref{2-3})  multiplied by an arbitrary
smooth function $M$ is again a conditional symmetry, we have examined also the operator $M(t,x_1,x_2,u)Q$
and  shown that no further results are  obtained.
\end{remark}

Now we  apply the $Q$-conditional symmetry  (\ref{5-1})  in reducing
the nonlinear BVP with the governing equation (\ref{5-0}) and
conditions
\begin{eqnarray}
&   x_2 = 0:  u^{-1/2} u_{ x_2}  = q_0 +{2} q_1 t, \label{5-6} \\
&   x_2 \rightarrow + \infty:  u_{ x_2} = 0. \label{5-7}
\end{eqnarray}
Operator  (\ref{5-1}) produces the ansatz
\begin{equation}\label{5-8}
u= ( t\phi(x_1,x_2) +h(x_1,x_2))^2,
  \end{equation}
where $\phi(x_1,x_2)$ is new unknown function. It can be noted that
ansatz  (\ref{5-8}) was proposed (and applied for finding exact
solutions) in \cite{king92} without knowledge of  symmetry
(\ref{5-1}). Substituting  (\ref{5-8})  into BVP (\ref{5-0}),
(\ref{5-6}) -- (\ref{5-7}) and taking into account (\ref{5-5}), we
arrive at two-dimensional problem for the nonlinear system of two
elliptic equations:

 \begin{equation}\label{5-9}
     \Delta  \phi= \phi h, \quad    \Delta h=h^2 ,
     \end{equation}
\begin{eqnarray}
&   x_2 = 0:   \phi_{ x_2}  = q_0, \quad  h_{ x_2}=q_1, \label{5-10} \\
&   x_2 \rightarrow + \infty:   \phi_{ x_2} = 0, \quad  h_{ x_2} = 0.\label{5-11}
\end{eqnarray}

\section{Some remarks about the domain geometry  }

 A natural question arises: how  do Lie and conditional invariance of BVPs depend on geometry of  the domain $\Omega$?
 Obviously the problem  essentially depends on the space dimensionality. For example, there are only
  three essentially different cases for  BVPs  with the (1+1)-dimensional evolution equations, namely:
   $\Omega$ is a finite interval,  a  semi-infinite interval  and $\Omega=\mathbb{R}$. Here we treated  (1+2)-dimensional  BVPs with $\Omega = \{(x_1, x_2): - \infty < x_1 < + \infty, \, x_2 > 0 \}$.
    In the general case,  the domain can be any open subset  $\Omega \subset \mathbb{R}^2$
     with a smooth boundary, i.e. one is formed by  differentiable (excepting possibly a finite number of points) curves.
    However, if one fixes a governing equation then the geometrical structure of  $\Omega$ may be predicted in advance if one is looking for Lie and conditional invariance of the relevant BVP.
   In the case of the governing   equation (\ref{0.1}),  all  possible Lie symmetries are presented in Table 1.
   Let us skip the critical case 3 because this  involves an infinite-dimensional algebra. The projection of all  MAIs  arising in cases 1,2 and 4 on the $(x_1, x_2)$-space gives the Lie algebra with basic operators
    \be\label{6-0}
  X_1 = \partial_{x_1}, \, X_2 = \partial_{x_2}, \,
J_{12}=x_1 \partial_{x_2}-x_2 \partial_{x_1}, \, D_{12} =  x_1 \partial_{x_1}+x_2 \partial_{x_2}, \ee
which is nothing else but the Euclidean algebra $AE(2)$  extended by the operator of scale  transformations. Now we realize that a non-trivial result can be obtained provided  $\Omega$ is invariant under  transformations generated by this algebra. For example, the case addressed above, namely   $\Omega = \{(x_1, x_2): - \infty < x_1 < + \infty, \, x_2 > 0 \}$, is invariant under $x_1$-translations and scale transformations generated by $D_{12}$; however, to note a simple such example, any triangle in the $(x_1, x_2)$-space does not admit any transformations generated by  (\ref{6-0}). Of course, the domain $\Omega = \mathbb{R}^2$ is invariant under the extended Euclidean algebra (\ref{6-0}); however,  this domain   is   appropriate to   initial value problems only (an interesting symmetry-based approach for solving such problems was proposed in \cite{goard-08}) while any boundary-value problem implies $\Omega \not= \mathbb{R}^2$.

It turns out that   all the domains admitting at least one-dimensional algebra can be described using the well-known results of classification of inequivalent (non-conjugate) subalgebras for the extended  Euclidean algebra, which are presented, for example, in
\cite{bar}.
The corresponding  list of subalgebras  can be divided on subalgebras of different dimensionality. We present only those of dimensionality  one and two because subalgebras of  higher dimensionality immediately lead to $\Omega = \mathbb{R}^2$.

 The one-dimensional subalgebras are
\[
 \left<X_1 \right>, \ \left<J_{12}\right>, \ \left<D_{12}\right>, \ \left<J_{12} + \beta D_{12}\right> (\beta>0),
\]
and  the  two-dimensional ones are
\[
 \left<X_1, X_2 \right>, \ \left<X_1, D_{12}\right>, \ \left<J_{12},  D_{12}\right>.
\]
Obviously, absolute invariants of each algebra can be easily calculated in explicit form (see, e.g., the relevant theory in \cite{olv}), hence,  we need only
to provide a geometrical interpretation for each algebra.
In the case of  the algebra  $\left<X_1 \right>$,  the absolute invariant is $x_2$, hence the domain $\Omega$ can be created by   lines of the form $x_2=const$. It means that there are only two
generic domains,  the
 strip $\Omega_1 = \{(x_1, x_2): - \infty < x_1 < + \infty, \, C_1< x_2 <C_2 \}$ and the  half-plane $\Omega_2 = \{(x_1, x_2): - \infty < x_1 < + \infty, \, x_2 > C_2 \}$
(hereafter $C_1$ and $C_2$   are  arbitrary consts). Any other domain admitting the $x_1$-translations  can be obtained via a combination of $\Omega_1$ and $\Omega_2$.

In the case of  the algebra  $\left<J_{12} \right>$,  the absolute invariant is $x_1^2+x_2^2$, hence the domain $\Omega$ can be created by
circles of the form $x_1^2+x_2^2=const$. This  means that there are only three
generic domains, the
interior of the circle $\Omega_1 = \{(x_1, x_2):   x_1^2+x_2^2<C_2 \}$,
the exterior of the circle $\Omega_2 = \{(x_1, x_2):   x_1^2+x_2^2>C_1>0 \}$ and   the
annulus   $\Omega_3 = \{(x_1, x_2):   0<C_1<x_1^2+x_2^2<C_2 \}$.

In the case of  the algebra  $\left<D_{12} \right>$,  the absolute invariant is $ \frac{x_1}{x_2} $,
hence the domain $\Omega$ can be created by  lines  of the form $x_1=const \ x_2$  and $x_2=0$.
It means that there are only two
generic domains, the
wedge $\Omega_1 = \{(x_1, x_2): C_1  x_2< x_1 < C_2  x_2  \}$
 and  the half-plane $\Omega_2 = \{(x_1, x_2): - \infty <
x_1 < + \infty, \, x_2 > 0 \}$.

Finally, in the case of  the one-dimensional algebra  $\left<J_{12} + \beta D_{12} \right>$,  the absolute invariant is $ \sqrt{x_1^2+x_2^2}\exp\left(-\beta  \arctan\frac{x_1}{x_2}\right) $, hence the domain $\Omega$ can be created by the curves  $\sqrt{x_1^2+x_2^2}=const \exp\left(\beta  \arctan\frac{x_1}{x_2}\right)$. In the polar coordinates $(r, \theta)$ such curves are the logarithmic  spirals $r=const   \exp(\beta \theta) $, and  one obtains   only the
generic domain  $\Omega = \{(r, \theta):  C_1 \exp(\beta \theta)  < r < C_2  \exp(\beta \theta)  \}$  ($0<C_1<C_2$),
 which is the  space between two spirals.

Examination of  two-dimensional subalgebras listed above do not   lead to any new domains. In fact, the first and the third produce  $\Omega = \mathbb{R}^2$ while the second leads only to the half-space $\Omega = \{(x_1, x_2): - \infty < x_1 < + \infty, \, x_2 > 0 \}$, which is a particular case of the domain $\Omega_2$ obtained above for the algebra  $\left<X_1 \right>$.

 The above considerations provide a symmetry based motivation for
investigating half-space problems, as we have done above. The other
domains just recorded  should be taken into account for further
application of the technique established above.

\section{Conclusions }

In this paper, a new  definition (see Definition 2) of conditional
invariance for BVPs is proposed.  It is shown that    Bluman's
definition \cite{bl-1974,bl-anco02} for Lie invariance of BVPs,
which is widely used  to find Lie symmetries  of  BVPs  with
standard boundary conditions,   follows as  a natural particular
case  from Definition 2.  Simple examples of  direct  applicability
of the definition  to nonlinear (1+1)-dimensional  BVPs,  leading to
both  known and new  results,   are   demonstrated.

 The main result of the paper consists in the  successful application  of the
  definition for  Lie  and conditional symmetry classification of  BVPs of the form   (\ref{0.1})--(\ref{0.3}).
   It turns out that  a wide range of possibilities arises for BVPs with the governing  (1+2)-dimensional nonlinear  heat equation if one looks for Lie symmetries. Depending on the form of the pair $(d(u); q(t))$  there are 15 different cases (see Table 2) in
    contrast to the 4 different cases only that  arise for the governing equation (\ref{0.1}).  In particular, we have proved that there is a special exponent, $k=-2$,  for the power diffusivity $u^k$ when BVP with non-vanishing flux on the  boundary admits additional Lie symmetry operators  compared to the case $k\not=-2$ (see cases 9 and 10 in Table 2). It should be stressed that the power $k=-2$ is not a special case for the governing   equation (\ref{0.1}) with $d(u)=u^k$ in two space dimensions, though in some respects this result reflects its exceptional status in one dimension. It is  worth noting that the well-known  critical power  $k=-1$, leading to an infinite-dimensional invariance algebra of  the (1+2)-dimensional nonlinear  heat equation, preserves its special status only in the case of  zero flux on the  boundary (see case 11  in Table 2 and Remark 4).

  In the case of conditional symmetry classification of the BVPs class  (\ref{0.1})--(\ref{0.3}), our result is modest because the governing equation (\ref{0.1}) admits a $Q$-conditional symmetry only  for the diffusivity $d(u)=u^{-1/2}$ \cite{arr-goard-br}. Hence we have  examined BVP (\ref{0.1})--(\ref{0.3}) with $d(u)=u^{-1/2}$ only and proved  that this problem is    conditionally  invariant under operator  (\ref{5-1}) provided restrictions (\ref{5-4}) hold.

In order to demonstrate the applicability of the symmetries derived, we used those for reducing the nonlinear BVP (\ref{0.1})--(\ref{0.3}) with power diffusivity $u^k$ and a constant non-zero flux (such problems are    common in applications and describe a wide range of phenomena depending on values of  $k$).
One motivation was  to investigate   the structure of the (1+1)-dimensional  problems obtained.
It turns out that  all the reduced problems
 (excepting  problem (\ref{4-16})--(\ref{4-18}))  are elliptic
 ones. Some of them are well-known (see  (\ref{4-23})--(\ref{4-25})), while others seem to be new and will be treated in a forthcoming paper.

Finally, we have described  a brief analysis of a  problem of
independent interest, which follows in a natural way from  the
theoretical considerations presented in Section 2. The problem can
be formulated as follows: how do  Lie and conditional invariance of
BVP depend on geometry of the domain, in which the given BVP is
defined? We have solved this problem for BVPs with the governing
equation (\ref{0.1}) and obtained an exhaustive  list of possible
domains preserving at least a one-dimensional  subalgebra of MAI of
equation (\ref{0.1}). It turns out that the geometrical
interpretation of the domains obtained is rather simple. However, we
foresee much more difficulties for BVPs in this regard with the
governing equations in  spaces of  higher dimensionality.

\section{Acknowledgements}
This research was supported by a Marie Curie International Incoming
Fellowship to the first author within the 7th European Community
Framework Programme.

\end{document}